\documentclass[prd,aps]{revtex4}

\usepackage{hyperref}
\usepackage{color}

\def\hhref#1{\href{http://arxiv.org/abs/#1}{arXiv:#1}} 

\begin{document}

\title{The Heisenberg-Euler Effective Action: 75 years on \footnote{Introductory review talk for the Heisenberg-Euler Session at {\href{http://benasque.org/2011qfext/}{QFEXT11}}, commemorating the 75th anniversary of the publication of the Heisenberg-Euler effective action:  W. Heisenberg and H. Euler, ``Consequences of Dirac's Theory of Positrons'', Zeit. f. Phys. {\bf 98}, 714 (1936).}}

\author{Gerald V. Dunne}

\affiliation{Physics Department, University of Connecticut, Storrs, CT 06269-3046, USA}

\begin{abstract}
On this 75th anniversary of the publication of the Heisenberg-Euler paper on the full non-perturbative one-loop effective action for quantum electrodynamics I review their paper and discuss some of the impact it has had on quantum field theory.
\end{abstract}

\maketitle

\section{Historical context}	

After the 1928 publication of Dirac's work on his relativistic theory of the electron  \cite{dirac1}, Heisenberg immediately appreciated the significance of the new  "hole theory" picture of the quantum vacuum of quantum electrodynamics (QED). Following some confusion, in 1931 Dirac associated the holes with positively charged electrons \cite{dirac2}:  
\begin{quote}
{\it A hole, if there were one, would be a new kind of particle, unknown to experimental physics, having the same mass and opposite charge to an electron.}
\end{quote}
\noindent
With the discovery of the positron in 1932, soon thereafter [but, interestingly, not 
immediately \cite{farmelo}],  Dirac proposed at the 1933 Solvay Conference that the negative energy solutions [holes] should be identified with the positron \cite{dirac3}: 
\begin{quote}
{\it Any state of negative energy which is not occupied represents a lack of uniformity and this must be shown by observation as a kind of hole. It is possible to assume that the positrons are these holes.}
\end{quote}
\noindent
Positron theory and QED was born, and Heisenberg began investigating positron theory in earnest, publishing two fundamental papers in 1934, formalizing the treatment of the quantum fluctuations inherent in this Dirac sea picture of the QED vacuum \cite{heisenberg1,heisenberg2}. It was soon realized that these quantum fluctuations would lead to quantum nonlinearities \cite{heisenberg2}:
\begin{quote}
{\it Halpern and Debye have already independently drawn attention to the fact that the Dirac theory of the positron leads to the scattering of light by light, even when the energy of the photons is not sufficient to create pairs.}
\end{quote}
\noindent
Halpern had published a brief note, without any details,  stating that light-light scattering could occur, and Heisenberg attributes Debye with suggesting, in private discussions,  the physical possibility of light-light scattering. Heisenberg set his student, Hans Euler, the task of studying light-light scattering using the density matrix formalism he had developed in \cite{heisenberg1,heisenberg2}. This work became Euler's PhD thesis \cite{eulerthesis} at Leipzig in 1936. A short paper in 1935, published  with another of Heisenberg's students, Bernhard Kockel, gave the results for the light-light scattering amplitude in the low frequency limit \cite{eulerkockel}. Soon after the Euler-Kockel paper, Akhieser published [with Landau and Pomeranchuk] a brief note on the light-light scattering amplitude in the high frequency limit, work that became Akhieser's thesis  under Landau's direction \cite{akhieser}.

The Euler-Kockel paper made it  clear that the quantum vacuum could be viewed as a  medium: 
\begin{quote}
{\it The connection between the quantities $\vec{B}$ and $\vec{D}$, on the one hand, and $\vec{E}$ and $\vec{H}$, on the other, is therefore nonlinear in this theory, since the scattering of light implies a deviation from the superposition principle.}
\end{quote}
\noindent 
They computed the leading quantum correction to the Maxwell Lagrangian:
\begin{eqnarray}
L=\frac{\vec{E}^2-\vec{B}^2}{2}+\frac{1}{90\pi}\frac{\hbar c}{e^2} \frac{1}{E_0^2}\left[ \left(\vec{E}^2-\vec{B}^2\right)^2+7\left(\vec{E}\cdot\vec{B}\right)^2\right]\quad ,
\label{quartic}
\end{eqnarray}
where they identified $E_0=e/(e^2/mc^2)^2$ as the (classical) "field strength at the electron radius". They interpreted this as vacuum polarization:
\begin{eqnarray}
\vec{D}&=&\vec{E}+\frac{1}{90\pi}\frac{\hbar c}{e^2} \frac{1}{E_0^2}\left[4 \left(\vec{E}^2-\vec{B}^2\right)\vec{E}-14\left(\vec{E}\cdot\vec{B}\right)\vec{B}\right]\nonumber\\
\vec{H}&=&\vec{B}+\frac{1}{90\pi}\frac{\hbar c}{e^2} \frac{1}{E_0^2}\left[4 \left(\vec{E}^2-\vec{B}^2\right)\vec{B}-14\left(\vec{E}\cdot\vec{B}\right)\vec{E}\right] \quad .
\label{polarization}
\end{eqnarray}
Euler and Kockel also computed the light-light scattering interaction cross-section, for mean wavelength $\lambda$:
\begin{eqnarray}
Q\sim \left(\frac{e^2}{\hbar c}\right)^4\left(\frac{\hbar}{mc}\right)^4\frac{1}{\lambda^2} \quad ,
\label{cross}
\end{eqnarray}
and concluded that the effect was tiny: 
\begin{quote}
{\it The experimental test of the deviation from the Maxwell theory is difficult since the noteworthy effects are extraordinarily small.}
\end{quote}

In modern language, Euler and Kockel studied QED vacuum polarization in the constant background field limit, obtaining the leading nonlinear corrections in powers of the field strengths. This was complementary to the work of Serber and Uehling \cite{serber},  at about the same time, who computed instead the corrections linear in the fields, but nonlinear in the space-time dependence of the background fields. Euler fully appreciated this distinction, in his thesis giving the general "effective field theory" form of the effective action as an expansion both in powers of the field strengths and in their derivatives. Euler and Kockel also commented on the formal similarity of the result (\ref{quartic}) to the work of Born and Infeld \cite{born}, who obtained similar nonlinear corrections to Maxwell theory, but from a  classical perspective.

In the 
Heisenberg-Euler paper \cite{he}, published in 1936, they
 extended the Euler-Kockel results in several significant ways. First, they obtained a closed-form expression for the full nonlinear correction to the Maxwell Lagrangian, a non-perturbative expression incorporating all orders in the (uniform) background electromagnetic field, presented in the abstract of their paper:
\begin{eqnarray}
{\mathcal L}&=&\frac{e^2}{h c}
\int_0^{\infty}\hskip -5pt \frac{d\eta}{\eta^3}
e^{-\eta}
\left\{i  \eta^2 (\vec{E}.\vec{B})\frac{\left[\cos\left(\frac{\eta}{ {\mathcal E}_c} \sqrt{\vec{E}^2-\vec{B}^2+2 i (\vec{E}.\vec{B})}\right)+{\rm c.c.}\right]}{\left[\cos\left(\frac{\eta}{ {\mathcal E}_c} \sqrt{\vec{E}^2-\vec{B}^2+2 i (\vec{E}.\vec{B})}\right)-{\rm c.c.}
\right]} 
+ {\mathcal E}_c^2+ \frac{\eta^2}{3} (\vec{B}^2-\vec{E}^2)
\right\} \quad .
\label{full}
\end{eqnarray}
Expanding this result to quartic order in a perturbative weak field expansion, one regains (\ref{quartic}), but  (\ref{full}) is  the full non-perturbative expression for the effective action. Significantly, 
 they expressed the result not in terms of the classical quantity $E_0$, as in (\ref{quartic}), but in terms of the critical field strength ${\mathcal E}_c=\alpha\, E_0$:
\begin{eqnarray}
{\mathcal E}_c=\frac{m^2 c^3}{e\hbar}\approx 10^{16} {\bf V}/{\rm cm}\quad ,
\label{critical}
\end{eqnarray}
which had already been identified by Sauter \cite{sauter} [who attributes the idea to Bohr] as the field strength scale at which one would expect Dirac sea electrons to tunnel into the continuum, producing electron-positron pairs from vacuum. Having the full non-perturbative result,  Heisenberg and Euler were able to identify this nontrivial prediction of positron theory:  the instability of the QED vacuum when subjected to a classical electric field, leading to the production of electron-positron pairs. Heisenberg and Euler understood that  background magnetic and electric fields lead to different physical effects \cite{he}: 
\begin{quote}
{\it In the presence of only a magnetic field, the stationary states can be divided into those of negative and positive energy. ... The situation is different in an electric field. ... This difficulty is physically related to the fact that in an electric field, pairs of positrons and electrons are created. The exact analysis of this problem was performed by Sauter.}
\end{quote}
\noindent Indeed, picturing this electron-positron pair production process as a tunneling process from the Dirac sea, they used Sauter's exact solution of the Dirac equation in a  constant electric 
field \cite{sauter} to estimate the rate of such a process as $\exp\left[-m^2 c^3 \pi/(\hbar e |E|)\right]$. Sauter had been a student in Munich, and after finding this exponential factor, stated \cite{sauter}: 
\begin{quote}
{\it This agrees with the conjecture of N. Bohr that was given in the introduction, that one first obtains the finite probability for the transition of an electron into the region of negative impulse when the potential ramp ${\mathcal E} h/mc$ over a distance of the Compton wavelength $h/mc$ has the order of magnitude of the rest energy ...  this case would correspond to around $10^{16} {\bf V}/{\rm cm}$.}
\end{quote}
\noindent This critical electric field value ${\mathcal E}_c$ in (\ref{critical}), is nowadays usually called the ``Schwinger critical field'' and serves as an estimate of the onset of the nonlinear QED region.

Heisenberg and Euler's computation was a {\it tour de force}, working with exact solutions of the Dirac equation in a constant electric and magnetic field background, combining Euler-Maclaurin summations over the Landau levels with integral representations of the parabolic cylinder functions, in order to derive the closed-form integral representation (\ref{full}) of the effective action. Their starting point was simple \cite{he}: 
\begin{quote}
{\it Due to relativistic invariance, the Lagrangian can only depend on the two invariants, ($\vec{E}^2-\vec{B}^2$) and $(\vec{E}\cdot\vec{B})$. The calculation of $U(\vec{E}, \vec{B})$ can be reduced to the question of how much energy density is associated with the matter fields in a background of constant fields $\vec{E}$ and $\vec{B}$.}
\end{quote}
\noindent Rewriting these invariants in terms of the eigenvalues of the field strength tensor, $a$ and $b$, where  $a^2-b^2=(\vec{E}^2-\vec{B}^2)/{\mathcal E}_c^2$, and $a\, b=(\vec{E}\cdot\vec{B})/{\mathcal E}_c^2$, Heisenberg and Euler also wrote the effective action  (\ref{full}) as \cite{he}
\begin{eqnarray}
{\mathcal L}&=&
4\pi^2 mc^2 \left(\frac{mc}{h}\right)^3
\int_0^{\infty}\frac{d\eta}{\eta^3}
\,e^{-\eta}
\left\{-a\, \eta\, {\rm ctg}(a \, \eta)\, b\, \eta\, {\rm cotanh}(b \, \eta)
+1
+\frac{\eta^2}{3} (b^2-a^2)
\right\}\, .
\label{full2}
\end{eqnarray}
They noted \cite{he}: 
\begin{quote}
{\it The integral (for $b=0$) around the pole $\eta=\pi/a$  has the value: $(-2i/\pi)4a^2mc^2(mc/h)^3\, e^{-\pi/a}$. This is the order of the terms which are associated with the pair creation in an electric field.}
\end{quote}

A third remarkable feature of the Heisenberg-Euler analysis is that they identified the physical significance of the subtraction terms in (\ref{full}, \ref{full2}). They noted that the first subtraction term corresponds to the subtraction of the (infinite) free-field effective action. 
They further realized that the other (logarithmically divergent) subtraction term had the functional form of the classical Maxwell action. Heisenberg was particularly interested in the physical significance of such logarithmically divergent terms, which now can be seen as an embryonic recognition of charge renormalization. 
Finally, the integral representation in  (\ref{full}, \ref{full2}) has the form that we nowadays refer to as the "proper-time form", as discussed in the next section.

Soon after the 1936 Heisenberg-Euler paper, Weisskopf presented a considerably simplified computation of the Heisenberg-Euler effective action, for both spinor and scalar QED, working directly from the spectrum of the Dirac [and Klein-Gordon] equation rather than from the eigenfunctions and density matrix. He also stated very clearly the new physical perspective \cite{weisskopf}: 
\begin{quote}
{\it The electromagnetic properties of the vacuum can be described by a field-dependent electric and magnetic polarizability of empty space, which leads, for example, to refraction of light in electric fields or to a scattering of light by light.}
\end{quote}

\section{Proper-time formulation: Feynman and Schwinger}

The next major developments in the subject came with work of Fock \cite{fock} and St\"uckelberg \cite{stueckelberg}, but this was not widely appreciated until the work of Feynman and Schwinger, who developed two different but complementary perspectives of the vacuum polarization problem in terms of  proper-time evolution. Feynman was trying to extend his path integral representation of non-relativistic quantum mechanics to the relativistic positron theory of Dirac. 
 Fock had formulated the Dirac equation in terms of proper-time evolution \cite{fock}, and St\"uckelberg \cite{stueckelberg} had proposed a physical picture in which positrons propagate backwards in real time, while proper-time evolves monotonically.  Feynman \cite{feynman1,feynmanappendix,feynman3} proposed to represent a quantum transition matrix element as a path integral over all paths in four dimensional space-time, with the paths parameterized by a fifth parameter \cite{feynmanappendix}:
 \begin{quote}
 {\it We try to represent the amplitude for a particle to get from one point to another as a sum over all trajectories of an amplitude $\exp[i \,S]$ where $S$ is the classical action for given trajectory. To maintain the relativistic invariance in evidence the idea suggests itself of describing a trajectory in space-time by giving the four variables $x_\mu(u)$ as functions of some fifth parameter $u$ ... (somewhat analogous to proper time).}
 \end{quote}
 \noindent Feynman noted that in such a construction there would be trajectories that appear to go backwards in time, but motivated by work of St\"uckelberg, had the brilliant idea to identify the forward-in-time paths with electrons and the backward-in-time paths with positrons. He showed that this led to a consistent path integral formulation of positron theory, incorporating all the pair-creation and annihilation processes of QED as twists and turns of space-time paths. Nambu put it most eloquently \cite{nambu}: \begin{quote}
 {\it The time itself loses sense as the indicator of the development  
of phenomena; there are particles which flow down as well as up the stream of time; the eventual creation and annihilation of pairs that may occur now and then, is no creation and annihilation, but only a change of directions of moving particles, from past to future, or from future to past; a virtual pair, which, according to the ordinary view, is foredoomed to exist only for a limited interval of time, may also be regarded as a single particle that is circulating round a closed orbit in the four-dimensional theatre; a real particle is then a particle whose orbit is not closed but reaches to infinity.}
\end{quote}

Feynman noted the important contributions of Fock, St\"uckelberg and Nambu. Nambu had extended Fock's proper-time approach  to the Klein-Gordon equation for scalar QED, and computed the path integral propagation amplitude for a constant background field.  In this approach, the Klein-Gordon equation [for scalar QED] appears as a Schr\"odinger-like equation:
\begin{eqnarray}
i\frac{\partial}{\partial u}\phi=-\frac{1}{2}\left(i\frac{\partial}{\partial x_\mu}-A_\mu\right)^2\phi
\equiv {\mathcal H}\,\phi \quad  .
\label{kg}
\end{eqnarray}
Studying a path integral representation of the corresponding  amplitude for evolution in $u$, the amplitude $\langle x|e^{-i\,{\mathcal H}\, u}|y\rangle$, Feynman arrived at what is now known as the world line representation of the [scalar] QED effective action:
\begin{eqnarray}
\Gamma[A]=-\int_0^\infty \frac{du}{u}e^{-m^2 u}\int d^4 x\int_{x(u)=x(0)=x} {\mathcal D}x\, e^{-S[x]} \quad ,
\label{wl}
\end{eqnarray}
where $S[x]$ is the classical action for a charged scalar particle to propagate on a space-time trajectory $x^\mu(\tau)$ for total proper time $u$:
\begin{eqnarray}
S[x]=\int_0^u d\tau\left(\frac{1}{2}\left(\frac{dx^\mu}{d\tau}\right)^2+e\, A_\mu \frac{dx^\mu}{d\tau}\right) \quad .
\label{action}
\end{eqnarray}
This worldline path integral representation (\ref{wl}) appeared in the appendix of one of Feynman's QED papers \cite{feynmanappendix}. Morette studied its mathematical basis \cite{morette}. However, Feynman's work on this first-quantized form of QED was largely unappreciated for many years. Interest was revived by the string theory work of Polyakov \cite{polyakov}, and the subsequent study of the field theory limit of string theory in the work of Bern and Kosower \cite{bernkosower}. Bern and Kosower showed that perturbative computations of scattering amplitudes in quantum field theory could be expressed in this first-quantized world line language, and in fact this leads to more efficient methods for certain multi-loop amplitudes \cite{bernkosower}. This idea has become a powerful modern method of computation in quantum field theory, from QCD to super-Yang-Mills, to supergravity. In fact, it also has a non-perturbative side, and this leads to a promising  way to compute the non-perturbative vacuum pair-production probability, extending the result of Sauter and Heisenberg-Euler to inhomogeneous background electric fields \cite{dunne-eli}. 

Soon after Feynman's work, Schwinger published a seminal paper \cite{schwinger1} re-formulating the results of Heisenberg and Euler in the new language of renormalized QED. Schwinger also viewed the QED processes as evolution in proper-time, but instead of a path integral method, he used first an operator solution of the proper-time evolution, and later developed a formalism based on Fredholm determinants \cite{schwinger2}. Schwinger's ``On Gauge Invariance and Vacuum Polarization'' paper presents a careful treatment of the Green's functions, studying renormalization and gauge invariance \cite{schwinger1}: 
\begin{quote}
{\it A renormalization of the field strength and charge, applied to the modified lagrange function for constant fields, yields a finite, gauge invariant result which implies nonlinear properties for the electromagnetic field in the vacuum.}
\end{quote}
\noindent He interprets the use of proper time as providing a gauge invariant regulator. This paper presents the exact result for the effective action for two special cases: first, the uniform background field treated by Heisenberg and Euler, and second the plane-wave background field, for which the Dirac equation had been solved by Volkov \cite{volkov}, and for which the effective action vanishes. Schwinger writes the effective Lagrangian in terms of the proper time evolution operator $U(s)$:
\begin{eqnarray}
{\mathcal L}(x)&=&\frac{1}{2}\, i\, \int_0^\infty \frac{ds}{s}\,\exp (-im^2 s)\, {\rm tr}(x|U(s)|x)\nonumber\\
U(s)&=&\exp(-i\, {\mathcal H}\, s)\qquad, \qquad  {\mathcal H}=\Pi_\mu^2-\frac{1}{2}e\sigma_{\mu\nu}F_{\mu\nu}\quad ,
\label{schwinger}
\end{eqnarray}
noting that \cite{schwinger1}
\begin{quote}
{\it $U(s)$ may be regarded as the operator describing the development of a system governed by the `hamiltonian', ${\mathcal H}$, in the `time' $s$, the matrix element [$(x'|U(s)|x'')$] of $U(s)$ being the transformation function from a state in which $x_\mu(s=0)$ has the value $x_\mu^{\prime\prime}$ to a state in which $x_\mu(s)$ has the value $x_\mu^\prime$.}
\end{quote}
\noindent 
Using results of Fock \cite{fock} and Nambu \cite{nambu} for $U(s)$ for a constant background field strength, Schwinger presents expressions for the effective Lagrangian in complete agreement with the expressions of Heisenberg-Euler \cite{he} and Weisskopf \cite{weisskopf}.

Furthermore, from the expression for a constant electric field ${\mathcal E}$,
\begin{eqnarray}
{\mathcal L}=\frac{1}{2}{\mathcal E}^2-\frac{1}{8\pi^2}\int_0^\infty \frac{ds}{s^3}\left[ e{\mathcal E}s\, {\rm cot}(e{\mathcal E}s)-1+\frac{1}{3}(e{\mathcal E}s)^2\right] \quad ,
\label{sch}
\end{eqnarray}
Schwinger extracted the full imaginary part, extending Heisenberg and Euler's derivation of the leading imaginary part, obtaining the instanton sum:
\begin{eqnarray}
2\, {\rm Im}\,
{\mathcal L}=\frac{\alpha^2}{\pi^2}\, {\mathcal E}^2\sum_{n=1}^\infty n^{-2}\, \exp\left(\frac{-n\, \pi\, m^2}{e\, {\mathcal E}}\right)
\quad .
\label{imag}
\end{eqnarray}
Soon afterwards, in number V of his series of six papers on QED, ``The Theory of Quantized Fields I-VI'',
Schwinger formally defines the effective action in terms of the determinant of the Dirac operator \cite{schwinger2}:
\begin{eqnarray}
\Gamma =-i\, \ln\, \det\left(1-e\gamma A\, G_+^{(0)}\right)\quad ,
\label{det}
\end{eqnarray}
where $G_+^{(0)}=1/(-i\partial\hskip -5pt / +m +i\epsilon)$ is the free Feynman propagator. Developing the theory of such Fredholm determinants, Schwinger shows that the proper-time representation leads naturally to an expression of the integral representation form (\ref{full}) found by Heisenberg and Euler in the constant background field case. This work puts the derivation of Heisenberg and Euler on a firmer theoretical and computational basis, and has been the inspiration for much of the subsequent development of quantum field theory in background fields. 

\section{Scientific legacy of Heisenberg and Euler}

The work of Heisenberg and Euler continues to have a profound impact even today, 75 years later. It is clear that they were well ahead of their time. Here I give a very abbreviated list of some developments that have come directly from their work. Of course it is not possible to be comprehensive in such a  short space. Further details can be found in \cite{greiner,dr-qed,dittrichgies,dk}.

\subsection{Light-light scattering}

The full light-light scattering amplitude was computed some years later by Karplus and Neuman \cite{karplus}. The effect is perturbatively very small and has not yet been directly observed. On the other hand, the related vacuum polarization effect of Delbr\"uck scattering has been seen \cite{delbruck}.

\subsection{Beta functions}

Weisskopf studied further \cite{weisskopf2} the logarithmic divergences that had been identified by Heisenberg and Euler, and which we now associate with charge renormalization. In fact, nowadays this idea provides a direct approach to compute the $\beta$ function for the running charge, using the external field scale rather than an external momentum scale. This provides an interesting perspective of vacuum polarization \cite{pagels,fujikawa,hansson} and   becomes a computationally powerful  method
at higher loops \cite{shifman}.

\subsection{Vacuum pair production}

While vacuum pair production was a definite, and quantitative, prediction in the Heisenberg-Euler \cite{he} paper [following the work of Sauter \cite{sauter}], the necessary electric field strength is so astronomical that it appeared out of experimental reach. 
After the discovery of lasers, the question was revisited by Br\'ezin \& Itzykson \cite{brezin} and Popov \& Marinov \cite{popov}, adapting to QED the seminal work of Keldysh \cite{keldysh} and collaborators in the theory of atomic ionization. Keldysh 
considered ionization not in a constant electric field but in a monochromatic time-dependent field $E(t)={\mathcal E}\,\cos(\omega\, t)$, defining a dimensionless "adiabaticity parameter" $\gamma$, being the ratio of the frequency $\omega$ to an inverse tunneling time. Remarkably, a WKB analysis of this ionization problem interpolates smoothly between the tunneling ionization regime where $\gamma\ll 1$, and the multi-photon ionization regime where $\gamma\gg 1$. Br\'ezin \& Itzykson \cite{brezin}, and Popov \& Marinov \cite{popov} applied a similar approach to the QED vacuum pair production problem, showing that the leading Sauter-Heisenberg-Euler exponential factor becomes [here $g(\gamma)$ is a simple  known function]
\begin{eqnarray}
\exp\left[-\frac{\pi m^2c^3}{e\hbar {\mathcal E}}g(\gamma)\right]\sim 
\begin{cases}
{\exp\left[-\frac{\pi m^2c^3}{e\hbar {\mathcal E}}\right]\qquad, \qquad \gamma\ll 1\cr
\left(\frac{e {\mathcal E}}{m\omega}\right)^{4mc^2/\hbar\omega}\qquad, \qquad \gamma\gg 1
}
\end{cases}
\end{eqnarray}
whose tunneling and multi-photon interpretations are self-evident. Since these fundamental works, there has been much theoretical work understanding this time-dependent pair production computation, using Bogoliubov transformations, quantum Vlasov equations, quantum kinetic theory, and semiclassical methods: for a recent review see \cite{dunne-eli}. The main outstanding question is how to increase the pair production rate by shaping the laser pulse appropriately. Unsolved problems concern understanding in a precise quantitative manner how the resulting predictions for the pair production rate would change when spatial focussing, back-reaction and cascading effects are included. These considerations have become more urgent, as recent work suggests that the peak electric field strength needed to observe some vacuum pair production may be lowered by one or two orders of magnitude \cite{dynamical,dipiazza,bulanov}, and this may be accessible in the not-too-distant future in several large-scale laser facilities \cite{ringwald,tajima,dunne-eli}. This, and related aspects of QED in ultra-strong laser fields, are  discussed further in Tom Heinzl's talk at this conference.

\subsection{Photon splitting, vacuum birefringence and axion searches}

In 1970 two groups \cite{bial,adler} computed the rate of photon splitting in a strong magnetic field, a process that can be described using  the Heisenberg-Euler effective action. The idea is to deduce the polarization tensor, from variation of the effective action, for the propagation of photons in a strong background field. Photon splitting has been experimentally measured now in \cite{splitting}. Further important progress came from the work of Tsai and Erber \cite{tsai}. Since the vacuum acts as a nonlinear medium, there can be both vacuum birefringence effects and also dichroism effects.
These results have become a paradigm in the growing field of precision tests of QED in strong external fields. The PVLAS [Polarizzazione del Vuoto con Laser] experiment is discussed in detail in Guido Zavattini's talk at this conference. The original PVLAS experiment reported an unexpectedly strong signal, which was not found in a revised experiment. However, this negative result had an unanticipated, but extremely  important, impact: it pushed people to study seriously the potential for axion searches using such strong-field experiments. There are now many such experiments [see the review by  H. Gies \cite{gies}], and this has become a {\it bona fide}  experimental field, complementary to astrophysical and accelerator searches.
Some recent theoretical results are discussed in Felix Karbstein's talk at this conference.

\subsection{Extensions of Heisenberg-Euler in QFT}

The nonabelian version of the Heisenberg-Euler effective action, for covariantly constant background gauge fields, was computed by Brown and Duff \cite{brownduff}, putting it in the language of Coleman-Weinberg effective potentials \cite{cw}. This led to further developments in nonabelian theories, for example for the gluon field in work of Matinyan and Savvidy \cite{matinyan}. Using the Fock-Schwinger gauge, in which $x_\mu A_\mu=0$, Novikov, Shifman, Vainshtein and Zakharov \cite{novikov} presented a simple systematic procedure to derive the leading terms in the large mass expansion for QCD [essentially the analogue of the Euler-Kockel quartic terms in (\ref{quartic})]. Their approach used the determinant formulation of Schwinger \cite{schwinger2}, and they showed how this relates to the operator product expansion and QCD sum rules. A covariant expansion based on the heat kernel definition of the determinant was developed by Barvinsky and Vilkovisky \cite{barvinsky}, a method directly applicable to both gauge and gravitational theories. This work emphasized the importance of non-local terms in the full effective action.

\subsection{Worldline approach to QFT}

Feynman's worldline formalism for the effective action has become a powerful new computational tool for quantum field theory, not just at one-loop but also at higher loops. Christian Schubert discusses some aspects of this in his talk. The  Bern-Kosower \cite{bernkosower,strassler,schubert} approach gives  a new way of doing perturbative scattering amplitude computations, leading to many simplifications and also some surprisingly simple results in maximally supersymmetric theories, including supergravity \cite{bern}.

\subsection{Effective actions in gravity and string theory}

Quantum field theory in curved space-time was heavily influenced by the work of Heisenberg and Euler, as is seen in the fundamental work of De Witt \cite{dewitt}, Parker \cite{parker}, Zeldovitch \& Starobinsky \cite{starobinsky}, Candelas \cite{candelas}, Davies \cite{davies}, Dowker \cite{dowker} and others [the work of Stuart Dowker is reviewed by Klaus Kirsten in this conference]. Much of this work is based on an adaptation of Schwinger's proper-time formulation of the effective action, from gauge theories to curved space-time. Indeed, the associated heat kernel expansion is often called the Schwinger-De Witt expansion. The study of particle creation in cosmological models, as developed by Parker \cite{parker} and Zeldovitch \& Starobinsky \cite{starobinsky}, is closely related to the vacuum pair production problem discussed by Heisenberg and Euler. The path-integral formulation of Bekenstein and Parker \cite{parker} starts with a generalization to curved space-time of the proper-time construction of Feynman and Schwinger. Recent progress for gravitational effective actions  is discussed at this conference in the talks by Ilya Shapiro and Alexei Starobinsky.
A particularly direct application of the Heisenberg-Euler effective action is in the work of Duff and Isham \cite{duffisham}, explaining the connection between self-duality, helicity and supersymmetry \cite{grisaru,kallosh}. This can be seen immediately in the SUSY QED combination of spinor and scalar effective actions, which when expressed in terms of the helicity fields $F_\pm$, one sees (for e.g.)  at the quartic level:
\begin{eqnarray}
{\mathcal L}_{\rm super}={\mathcal L}_{\rm spinor}+2{\mathcal L}_{\rm scalar}=\frac{\alpha^2}{12m^4}F_+^2\, F_-^2
\quad .
\label{super}
\end{eqnarray}
This clearly vanishes when either of $F_\pm$ vanishes. Remarkably, a similar feature has been seen in the Heisenberg-Euler effective action, even at the two-loop level, for super-Yang-Mills theories \cite{kuzenko}. The effective action formalism, based on the determinantal form, was extended to string theory by Fradkin and Tseytlin \cite{fradkin}, and has become a corner-stone of string theory and gauge-gravity dualities.

\subsection{Heisenberg-Euler effective action and zeta functions}

In 1977 Hawking  published an influential paper \cite{hawking} that introduces a definition of the effective action in terms of the zeta function of the relevant operator [Klein-Gordon, Dirac, ...]. This was based on mathematical work of Seeley. For an operator $D$ with spectrum $\lambda_n$, we formally define the zeta function as
$\zeta(s)=\sum_n \lambda_n^{-s}$, so that 
\begin{eqnarray}
\ln \det D=-\zeta^\prime(0)+\ln (\pi\mu^2/4)\zeta(0)\quad ,
\label{zeta}
\end{eqnarray}
where $\mu$ is a renormalization scale. Shortly after, Dittrich \cite{dittrich} showed that the Heisenberg-Euler effective action (\ref{full}) could be computed straightforwardly in this zeta function language, with the relevant zeta function being the Hurwitz zeta function, a generalization of the familiar Riemann zeta function \cite{ww}, essentially because the Landau-level-type  spectrum is linear in an integer index. These ideas have been extended in many ways \cite{elizaldebook,klaus}, and a mathematically elegant way to express the results is in terms of the Barnes multiple gamma function \cite{ruijsenaars}.

\subsection{Strong field QED: magnetic catalysis \& the chiral magnetic effect}

As discussed by Maxim Chernodub at this conference, there has been a great deal of recent work analyzing the effect of strong electromagnetic fields on the strong interactions of QCD. This has been spurred by astrophysical considerations, with ultra-strong magnetic fields known to be present in astrophysical objects such as neutron stars, as well as recent results [both experimental and theoretical] from heavy ion collisions at RHIC. In such collisions, huge magnetic fields are generated by the ions, and these can lead to surprisingly large effects. For example, Kharzeev \cite{dima} has proposed that the observed  asymmetry of particle correlations in such collisions \cite{star} may be explained using the ``chiral magnetic effect'', in which charged chiral fermions, associated with QCD vacuum fluctuations, are accelerated apart in a  strong magnetic field due to the lowest Landau level projection onto definite spin. This is technically  close to the magnetic catalysis mechanism of Gusynin, Miransky and Shovkovy \cite{miransky}, also based on the strong magnetic field limit of a Heisenberg-Euler-type computation, which makes important predictions for dynamical symmetry breaking in the presence of strong fields, and also for quantum transport and the quantum Hall effect in graphene. 

\section*{Acknowledgments}

I thank the organizers, especially Manuel Asorey and Michael Bordag, for an excellent conference. I also thank Walter Dittrich for his input on matters of both history and physics, and D. H. Delphenich for providing unpublished translations of several relevant papers. I acknowledge support from the US DOE through grant DE-FG02-92ER40716.

\end{document}